%% file: ms.tex
\documentclass[iop]{emulateapj}
\slugcomment{{\sc Accepted to ApJ:} February 24, 2012} 
\usepackage{epsfig}
\shorttitle{VLM Binary Rotational Velocities}
\shortauthors{Konopacky et al.}
\begin{document}

\title{Rotational Velocities of Individual Components in Very Low Mass Binaries}
\author{Q.M. Konopacky\altaffilmark{1,2},
  A.M. Ghez\altaffilmark{3}, D.C. Fabrycky\altaffilmark{4}, B.A. Macintosh\altaffilmark{1},
   R.J. White\altaffilmark{5},
  T.S. Barman\altaffilmark{6}, E.L. Rice\altaffilmark{7,8},
  G. Hallinan\altaffilmark{9}, G. Duch\^{e}ne\altaffilmark{10,11}}
\altaffiltext{1}{Lawrence Livermore National
  Laboratory, 7000 East Avenue, Livermore, CA 94550;
  macintosh1@llnl.gov}
\altaffiltext{2}{Current Address: Dunlap Institute for
  Astronomy and Astrophysics, University of Toronto, 50
  St. George Street, Toronto M5S 3H4, Ontario, Canada; Dunlap
  Fellow; konopacky@di.utoronto.ca}
\altaffiltext{3}{UCLA Division of Astronomy and Astrophysics, Los Angeles, CA 90095-1562; ghez@astro.ucla.edu}
\altaffiltext{4}{Department of Astronomy and Astrophysics,
  University of California, Santa Cruz, CA 95064, USA; Hubble
  Fellow; fabrycky@ucolick.org}
\altaffiltext{5}{Department of Physics and Astronomy, Georgia
  State University, Atlanta, GA 30303; white@chara.gsu.edu}
\altaffiltext{6}{Lowell Observatory, 1400 W. Mars Hill Rd.,
  Flagstaff, AZ 86001; barman@lowell.edu}
\altaffiltext{7}{American Museum of Natural History, Central
  Park West at 79th Street, New York, NY 10024-5192; erice@amnh.org}
\altaffiltext{8}{Current Address: College of Staten Island,
  City University of New York, 2800 Victory Blvd, Staten Island, NY 10314}
\altaffiltext{9}{Department of Astrophysics, California
  Institute of Technology, MC 249-17, Pasadena, CA 91125; gh@astro.caltech.edu}
\altaffiltext{10}{Astronomy Department, UC Berkeley, Hearst
  Field Annex B-20, CA 94720-3411, USA; gduchene@berkeley.edu}
\altaffiltext{11}{UJF-Grenoble 1 / CNRS-INSU, Institut de
  Plan\'etologie et d'Astrophysique 
de Grenoble (IPAG) UMR 5274, Grenoble, F-38041, France}

\begin{abstract}
We present rotational velocities for individual components
of eleven very low mass (VLM) binaries with spectral types
between M7 and L7.5.  These results are based on observations
taken with the near-infrared spectrograph, NIRSPEC, and the
Keck II laser guide star adaptive optics (LGS AO) system.  We find that
the observed sources tend to be rapid rotators ($v\,$sin$\,i$ $>$ 10 km
s$^{-1}$), consistent with previous seeing-limited measurements of VLM
objects.  The two sources with the largest $v\,$sin$\,i$, LP 349-25B and
HD 130948C, are rotating at $\sim$30$\%$ of 
their break up speed, and are among the most rapidly rotating
VLM objects known.  Furthermore, five binary systems, all with
orbital semi-major axes 
$\lesssim$3.5 AU, have component $v\,$sin$\,i$ values
that differ by greater than 3$\sigma$.  To bring the binary components with  
discrepant rotational velocities into agreement would require
the rotational axes to be inclined with respect to each other,
and that at least one  
component is inclined with respect to the
orbital plane.  Alternatively, each component could be rotating
at a different rate, even though they have similar spectral
types.  Both differing rotational velocities and inclinations
have implications for binary star formation and
evolution.   We also investigate possible dynamical
evolution in the triple system HD 130948A-BC.  The close binary
brown dwarfs B and C have
significantly different $v\,$sin$\,i$ values.  We
demonstrate that components B and C could have been torqued into misalignment by the
primary star, A, via orbital precession.  Such a scenario can also
be applied to another triple system in our sample, GJ
569A-Bab.  Interactions such as 
these may play an important role in the dynamical evolution of
very low mass binaries.  Finally, we note that two of the binaries with large
differences in component $v\,$sin$\,i$, LP 349-25AB and 2MASS
0746+20AB, are also known radio sources. 

\end{abstract}

\section{Introduction}

\input{tab1}

Rotational velocity is an important diagnostic parameter for
stellar objects, offering a window into the angular momentum
evolution of a given source.  A star's rotation
can provide important clues to its formation and can
furnish diagnostics of its interior structure and evolution.
For instance, measurements of rotational velocity have 
been shown to correlate strongly with stellar activity,
possibly driving the magnetic dynamo responsible for generating this
activity \citep{browning08}.  In
addition, rotational velocities have been shown to correlate with the age of a
system, offering a tool for estimating stellar ages \citep{delfosse98}.  

The rotational behavior of very low mass (VLM) stars and brown
dwarfs has been studied by a number of authors in recent
years \citep{mohanty03,bailer04,zapo06,reiners08,reiners10,blake10}.
It has been shown that the brown dwarfs tend to 
be rapid rotators, and that the minimum rotation rate is a
function of spectral type (i.e., \citealt{zapo06,reiners08}).  It has
also been determined that the rotational velocities of brown dwarfs
correlate with age, with VLM objects having very long spindown
timescales, and that their rotational evolution is 
probably dominated primarily by magnetic braking (e.g.,
\citealt{reiners08,scholz09,scholz11}).  However, 
it appears that the activity-rotation
relationship that is very strong amongst M dwarfs tends to
break down at these low masses \citep{mohanty03}.  In spite of
this, activity in the form of radio emission has been
observed in a number of VLM systems (e.g,
\citealt{berger06,osten06,hallinan08}).  In addition, it has
been shown that this drop in activity does not seem to be
due to a reduction in magnetic field strengths in late M
dwarfs, but might instead be due to their reduced temperature and hence
reduced fractional ionization of their atmospheres (e.g., \citealt{reinersb07,hallinan06,hallinan08}).  

The majority of previous studies have been performed with
seeing-limited observations, and most sources targeted are
thought to be single.  Known binaries have been included in
various samples, and their rotational velocities have been
derived from the combined light of both 
components.  The rotational velocities 
of individual binary components can potentially provide a
unique look at the rotational evolution of VLM objects.  If any
differences are seen between the velocities of the binary
components, it could have implications for the way in which
these binaries formed, their early accretion history, or the
operation of magnetic braking as a function of mass.  For instance,
\citet{reiners07} found that the components of the triple
system LHS 1070 (spectral types M5.5, M9, and M9) had differing $v\,$sin$\,i$, with the
higher mass component rotating about a factor of two more slowly than
the two lower mass objects.  This allowed the authors to put
constraints on the form of rotational braking in the VLM regime.  
Further,  \citet{gomez09}
found that the components of the young eclipsing binary
brown dwarf 2MASS0535-05AB (spectral types M6.5) have different rotational periods,
with the primary component rotating more rapidly than the
secondary.   

If the orbits of these binaries are known, rotational
velocities provide a way to test the assumption that spin axes
are generally perpendicular to the orbital plane.  Work by
\citet{hale94} found that, in general, binaries with 
separations $\lesssim$30-40 AU should have spin axes
perpendicular to their orbital plane.  However, some very close
(semi-major axis $\lesssim$ 0.3 AU) 
binaries such as DI Herculis \citep{albrecht09} have been shown
to have extremely misaligned axes.  In contrast, more recent
work on a very similar binary system (NY Cep, \citealt{albrecht11})
has revealed no such misalignment, implying that the cause of
the inclined axes is non-universal.  It is also
important to explore systems with wider separations
that may not be subject to the same extreme tidal interactions
as very close binaries.  This has been done in the case of
some T Tauri binary systems, which have shown both slight and
substantial planar misalignment via observations of disk
orientation \citep{jensen04, monin06}.  Probing the rotational
evolution of intermediate separation
binaries ($\thicksim$1 - 10 AU) in the substellar regime will
determine whether such trends hold at the lowest masses, with
interesting implications for the formation and evolution of
all types of binary stars.

In this paper, we present projected rotational velocity measurements for the
components of a sample of tight, visual VLM binaries.  The measurements of
these spatially-resolved velocities are enabled by the
W.M. Keck Observatory laser guide star adaptive optics (LGS
AO) system, which provides high spatial resolution
observations of optically faint targets \citep{wiz06}.  This study is the
first to systematically examine the rotational velocities of
individual VLM objects that reside in binary systems.  In
Section \ref{data}, we describe our observations and our
method for extracting rotational velocities from high
resolution spectra.  In Section \ref{disc}, we compare our
measurements to those of single VLM objects, and discuss the
implications of our measurements for theories of binary star
formation and evolution.  We also discuss our results in the
context of previously measured radio emission from two of our
sources.  We summarize our findings in Section \ref{conc}.

\section{Data and Analysis}\label{data}

\subsection{Sample}\label{samp}

Our sample is comprised of eleven VLM binaries that were
targeted as part of an ongoing program to measure their
dynamical masses.  These objects have been observed both
astrometrically and spectroscopically since 2006, and initial
estimates of their orbital properties have been obtained from
this dataset \citep{kono10}.  Their spectral types range from M7.5 to L7.5, and
their separations range from 0$\farcs$07 to 0$\farcs$35.
Because we are able to spatially resolve the components before
obtaining high resolution spectroscopy (see section
\ref{obs}), our total sample consists of 22 VLM objects.
Table \ref{tab:sample} summarizes the targets in our sample.

\input{tab2}

\subsection{Observations}\label{obs}
The eleven binaries were observed using the
NIR spectrograph NIRSPEC on Keck II 10 m (McLean et al. 2000) in
conjunction with the facility LGS AO system (NIRSPAO).  These
observations, taken between 2006 December and 2011 June, are
described in detail in Konopacky et al. (2010).  Briefly, we
used the instrument in its high spectral resolution mode,
selecting a slit 0$\farcs$041 in width and 2$\farcs$26 in length in AO
mode.  We observed in the K band in order to obtain data in
the CO bandhead region (2.291 - 2.325 $\mu$m, order 33).  Due to
the dense population of lines in this region, our analysis for
this work was done only in order 33, although the
cross-dispersed data ranged from 2.044 - 2.382 $\mu$m.

The camera was rotated such that both components of each binary fell
simultaneously on the high resolution slit, which is
at an angle of 105.9$^{o}$ with respect to vertical.  Typical observations
consisted of four spectra of both components, each with 900-1800 second
integration times, taken in an ABBA dither pattern along
the length of the slit.  
On average, we achieved Strehl ratios between 10-40$\%$ at K band,
resulting in PSF core full width half maxima of $\thicksim$0$\farcs$05-0$\farcs$08.  As
discussed in \citet{kono10}, this performance allowed us, in general, to
obtain resolved spectra for binaries separated by at least $\thicksim$0$\farcs$06.   

Table \ref{tab:obslog} gives the log of our
spectroscopic observations, 
listing the targets observed, the date of observation, the
number of spectra, the integration time for each
spectrum, and the average SNR achieved.  Because the spectra
are dominated by molecular features (see Figure \ref{fig:spec_all}), we have
chosen to estimate effective SNR by calculating the average number
of electrons per pixel in an extracted spectrum and then assuming
Poissonian noise on that average.  We have verified that these
estimates are roughly correct by calculating the SNR on small
regions of our most rapidly rotating source, where all
features are fairly smoothed out, and by using the
properties of the NIRSPEC detector under the
assumption that we are background limited at K
band\footnote{http://www2.keck.hawaii.edu/inst/nirspec/sens.html}.
Each target observation was accompanied by the
observation of a nearby A0V star to measure the telluric
absorption. 

\subsection{Data Reduction}\label{red}

As described in \citet{kono10}, the basic reduction
of the NIRSPAO spectra was performed with 
REDSPEC, a software package designed for
NIRSPEC\footnote{http://www2.keck.hawaii.edu/inst/nirspec/redspec/index.html}.
Object frames are reduced by subtracting opposing
nods to remove sky and dark backgrounds, dividing by a flat
field, and correcting for bad pixels.  Order 33 was spatially
rectified by fitting the trace of each nod of A0 calibrators with third order
polynomials, and then applying the results of those fits
across the image.   As these systems are fairly tight
binaries, cross-contamination can be an issue when extracting
the spectra.  We use a Gaussian extraction method to extract
the spectra, fitting the trace with a variable FWHM in cases
where the separation was greater than 7 pixels and a fixed
FWHM if less than 7 pixels.  We subtract the results of this
fit for one component before extracting the other (see
\citealt{kono10} for more details).  

\begin{figure}
%\epsscale{1.0}
\epsfig{file=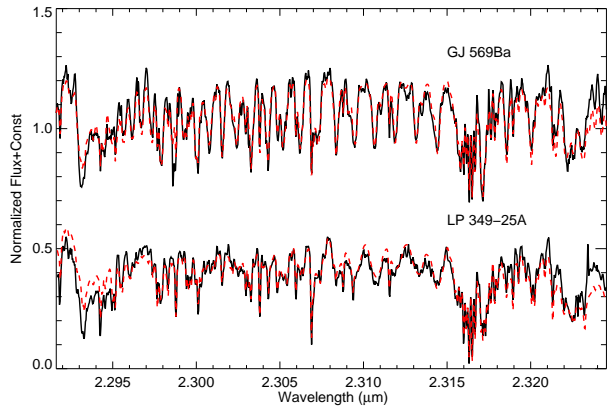,angle=90,width=\linewidth}
\caption{Example fits for two objects in our sample, GJ 569Ba
  and LP 349-25A.  The spectra have been normalized and separated on the
  y-axis by a constant vertical shift for visual clarity.  The black solid lines show the actual
  NIRSPAO data (order 33), which is not corrected for telluric
  absorption.  The red dashed lines show the
  fitted template, which is a combination of a theoretical
  PHOENIX spectrum and a telluric absorption model.  The
  $v\,$sin$\,i$ values
  measured from these particular fits are 18 km s$^{-1}$ for GJ
  569Ba and 56 km s$^{-1}$ for LP 349-25A.  For additional
  examples of fits for all sources in our sample, see Figure 4
of \citet{kono10}.}
\label{fig:fit_plot}
\end{figure}

\subsection{Determination of \textit{v}sin\textit{i}}

The data set presented here is the same set presented in
\citet{kono10}, except we now include three new epochs of data from December 2009,
June 2010, and June 2011.  In \citet{kono10} we were primarily interested in the
radial velocities, and hence orbital solutions, that could be derived from these spectra.
Here we reanalyze these data having implemented two changes to
our analysis in order to properly determine $v\,$sin$\,i$
\citep{bailey12}.  First, we now use the telluric lines
present in the spectra of the A0V 
calibrator stars to measure the instrumental profile for NIRSPAO.
Secondly, we are now performing the convolution 
of theoretical templates with a Gaussian kernel after putting
all spectra on a log-linear scale.  This makes the resolution
constant across the entire spectral range, providing a more
accurate measure of $v\,$sin$\,i$.  We describe our analysis in more
detail below.  Note that this reanalysis does not
substantially impact our radial velocity estimates, which will
be presented in a future paper.

It has been demonstrated that the CO bandhead line depths and
widths are primarily a function of temperature and the
projected rotational velocity ($v\,$sin$\,i$) for VLM objects,
respectively, with an additional 
moderate dependence on surface gravity \citep{blake07} and
metallicity.  With some knowledge 
of the temperature of a given object and an allowance for
unknown surface gravity and metallicity, $v\,$sin$\,i$
measurements can be estimated 
from our extracted spectra.  

Our extracted spectra are not corrected for telluric
absorption because these features provide a stable
reference for absolute wavelength calibration.  Using
features that are naturally present in all spectra
also allows us to accurately calibrate the instrumental profile
without the need to observe additional template sources.  We
therefore model each spectrum  as a combination of a KPNO/FTS
telluric spectrum  
\citep{livingston91} and a synthetically generated
spectrum derived from the PHOENIX atmosphere models
\citep{hauschildt99}.  The model spectrum is parameterized
to account for the wavelength solution, continuum normalization, 
instrumental profile (assumed to be Gaussian), $v\,$sin$\,i$, and
radial velocity.  As mentioned above, the
instrumental profile is determined using our A0V calibrator stars, which
by design are a clean measure of the actual telluric
spectrum.  We hold the instrumental profile fixed while
fitting our actual target spectra.  The average
resolution of our NIRSPAO data is $\sim$10 km s$^{-1}$.  The best-fit model is
determined by minimizing the variance-weighted reduced
$\chi^2$ of the difference between the model and the extracted
spectrum, once this difference has been Fourier filtered to
remove the fringing present in NIRSPEC K-band  
spectra (see \citealt{bailey12} for more details).  This model
therefore provides our $v\,$sin$\,i$ estimates.

\begin{figure*}
\epsscale{0.8}
\plotone{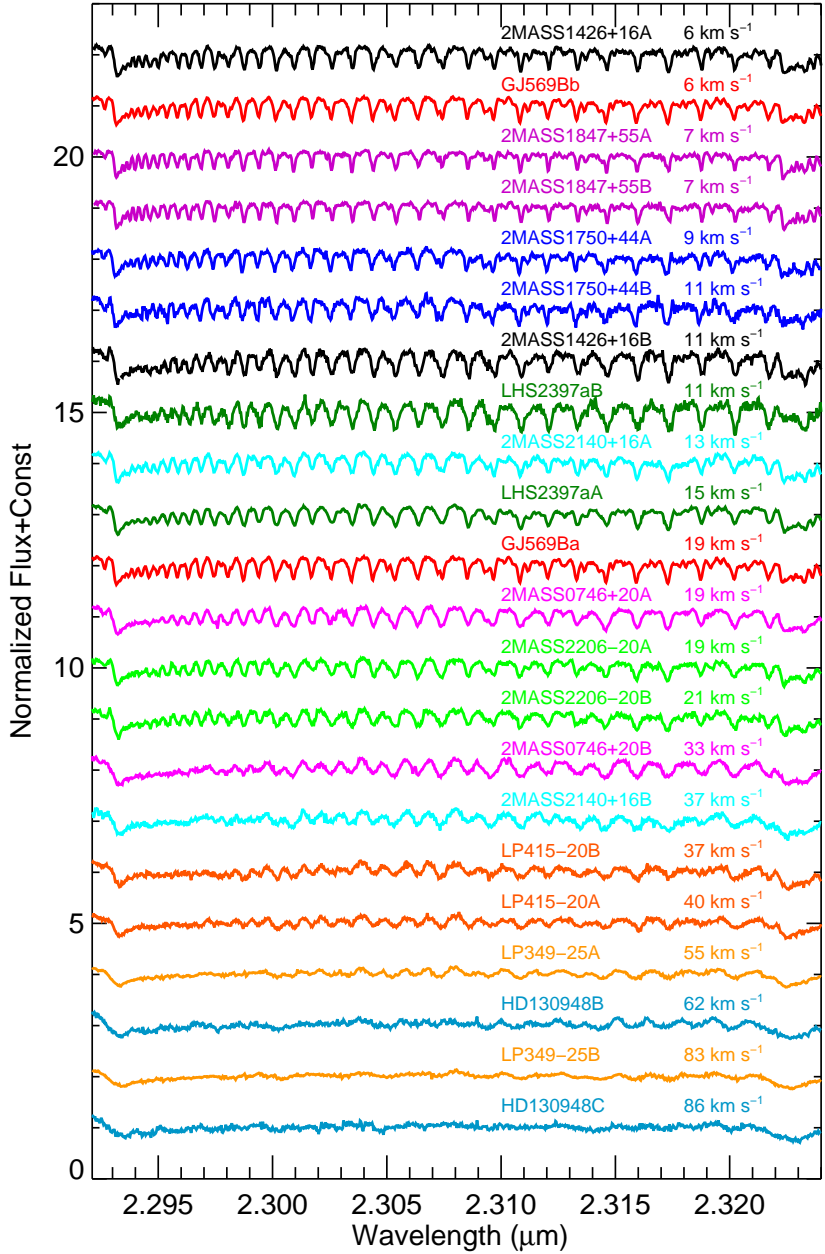}
\caption{Single epoch examples of spectra for all objects in
  our sample (NIRSPAO order 33).  All spectra have been
  normalized to the continuum and shifted to the same radial
  velocity.  The spectra have been separated on the y-axis by
  a constant vertical shift, such that the absolute value of
  the flux is arbitrary.  Sources are arranged in order of increasing
  rotational velocity and color-coded such that
  components in the same system have matching colors.  Increasing rotational
  velocity dramatically impacts the morphology of the CO
  bandhead, which falls near the end of the K band at
  $\thicksim$2.3$\mu$m.}
\label{fig:spec_all}
\end{figure*}

\input{tab3}

Each PHOENIX template is generated at a fixed temperature and
surface gravity.  The use of theoretical rather than observed
templates has the advantage of introducing less template
mismatch biases.  Our main templates for each source have a
temperature as measured in \citet{kono10}, a log(g) of 5.0,
and solar metallicity.  Figure \ref{fig:fit_plot} shows example fits
for two objects in our sample, one moderate rotator (GJ 569Ba) and
one rapid rotator (LP 349-25A) that have different
temperatures.  The figure demonstrates the location and morphology of the
telluric features that are used for estimating the
instrumental profile and wavelength solution.  Figure 4 of
\citet{kono10} shows example fits for all objects in our
sample - we refer the reader to this work for further visual
evaluation of our fitting technique. 

Statistical uncertainties are assigned by fitting each
individual spectrum separately and taking the RMS of the
values derived for each case.  We also need to account for
systematic uncertainties due to both the temperature and
surface gravity dependence of our spectra.  We
fit each spectrum with templates spanning $\pm$300 K in
temperature and ranging from 3.0 - 4.5 dex in log(g).  We also
explored varying metallicity, using
templates between $\pm$0.25 dex of solar to fit our spectra based
on metallicity measurements of low mass objects in the solar
neighborhood (e.g., \citealt{johnson09,schlaufman10,rojas10}).  We then use the
spread in these values around our best-fit value as our
systematic uncertainty, and add these in quadrature with our statistical
uncertainties.  We find on average that log(g)
uncertainties add a 3 km s$^{-1}$ uncertainty to the
$v\,$sin$\,i$, while temperature and metallicity uncertainties contribute an
additional 1 km s$^{-1}$ each, with lower values of
temperature and metallicity yielding lower $v\,$sin$\,i$.  The
$v\,$sin$\,i$ measured for each 
source at each epoch, along with the weighted average of all
epochs, is given in Table \ref{tab:vsini} (all sources were
observed at least two times).

In order to confirm that our method returns the correct
$v\,$sin$\,i$ values, we obtained NIRSPAO observations of two
previously-measured M type stars.  These objects, GL 1245A (M5.5V) and
G188-38 (M4V), where targeted by several studies in the
optical.  \citet{mohanty03} measured projected rotational velocities
of 22.5 $\pm$ 3.7 km s$^{-1}$ for GL 1245A and 29.4 $\pm$ 6.2
km s$^{-1}$ for G188-38.  Other measurements for G188-38
include 36.5 $\pm$ 0.3 km s$^{-1}$ by \citet{donati06} and
29.4 $\pm$ 1.4 km s$^{-1}$ by \citet{delfosse98}.  We
performed an identical analysis 
to that of our brown dwarf sample on these two mid-M stars,
only using a higher temperature template that is more appropriate for
these objects.  We derived $v\,$sin$\,i$s of 19 $\pm$ 3 km
s$^{-1}$ for GL 1245A and 34 $\pm$ 3 km s$^{-1}$ for G188-38,
consistent with all the values from the literature.

As an additional test, we can use ``slow rotators'' in our
sample as templates, artificially spinning them up to estimate
the $v\,$sin$\,i$ of other objects in our sample.  Although
this method suffers from template 
mismatch that is remedied by the use of theoretical atmospheres,
it offers a further confirmation of our technique.  We perform
this analysis on one epoch of data for each binary, taken
either in 2007 or 2008.  We use the source with the lowest
measured $v\,$sin$\,i$, 2MASS1847+55A, as our template for all
other objects.  This template, corrected for telluric
absorption using our observed A0V standards, was ``spun up'' to produce an
artificial grid of spectra with $v\,$sin$\,i$ between 5 and
100 km s$^{-1}$.  The grid was then cross-correlated with each
object's spectrum, also corrected for telluric absorption, and
we determined the value of $v\,$sin$\,i$ that provided the
maximum correlation.  In all cases, the best value for $v\,$sin$\,i$
found with this technique is within the uncertainties of the
values given in Table \ref{tab:vsini}.

Using these two independent checks, we are confident
that our methodology is sound and that we are incorporating
the necessary uncertainties via our usage of multiple
temperature and log(g) templates.  We do caution, however,
that objects with particularly high values of $v\,$sin$\,i$
($>$40 km s$^{-1}$, four objects in our sample), though
undoubtedly rapid rotators, might 
be subject to additional systematic uncertainties not fully
accounted for in our analysis due to greater sensitivity to
properties associated with the instrument and technique, and
this uncertainty may not be captured in the averaged values in
Table \ref{tab:vsini}.    

We also estimate the lowest measureable value of $v\,$sin$\,i$
in our spectra.  To do this, we took our PHOENIX
templates and broadened them first to the correct instrumental profile
and then to different values of $v\,$sin$\,i$.  We also
injected random Gaussian noise such that the templates would have SNR $\sim$
55, which is the average effective SNR per pixel for our data.  We fit these
spectra using the method described above.  We find that
the limiting value for which we could accurately measure
$v\,$sin$\,i$ is 3 km s$^{-1}$.  Note that we are able to
measure $v\,$sin$\,i$ below the instrinsic resolution of our
NIRSPAO data due to our accurate measurement of the instrumental
profile and theoretical templates that closely match our actual
spectra.

Figure \ref{fig:spec_all} shows example spectra for all
sources in our sample.  These spectra, which we have corrected
for telluric absorption for plotting purposes, are arranged in
order of increasing $v\,$sin$\,i$, demonstrating the effect
of rotational velocity on CO bandhead morphology.  

\vspace{1.7in}
  
\section{Discussion}\label{disc}

This study represents the first measurement of component
rotational velocities for a large sample of VLM binaries.
The values presented in Table \ref{tab:vsini} show that
$\thicksim$80$\%$ of our sample are rapid rotators ($v\,$sin$\,i$ $\gtrsim$ 
10 km s$^{-1}$), and two sources, LP 349-25B and HD130948C, are among the
fastest rotating VLM objects ever observed.  In this section,
we discuss the implications of these measurements.  

\subsection{Comparison to $v\,$sin$\,i$ Measurements in the Literature}

\begin{figure}
%\epsscale{1.0}
\epsfig{file=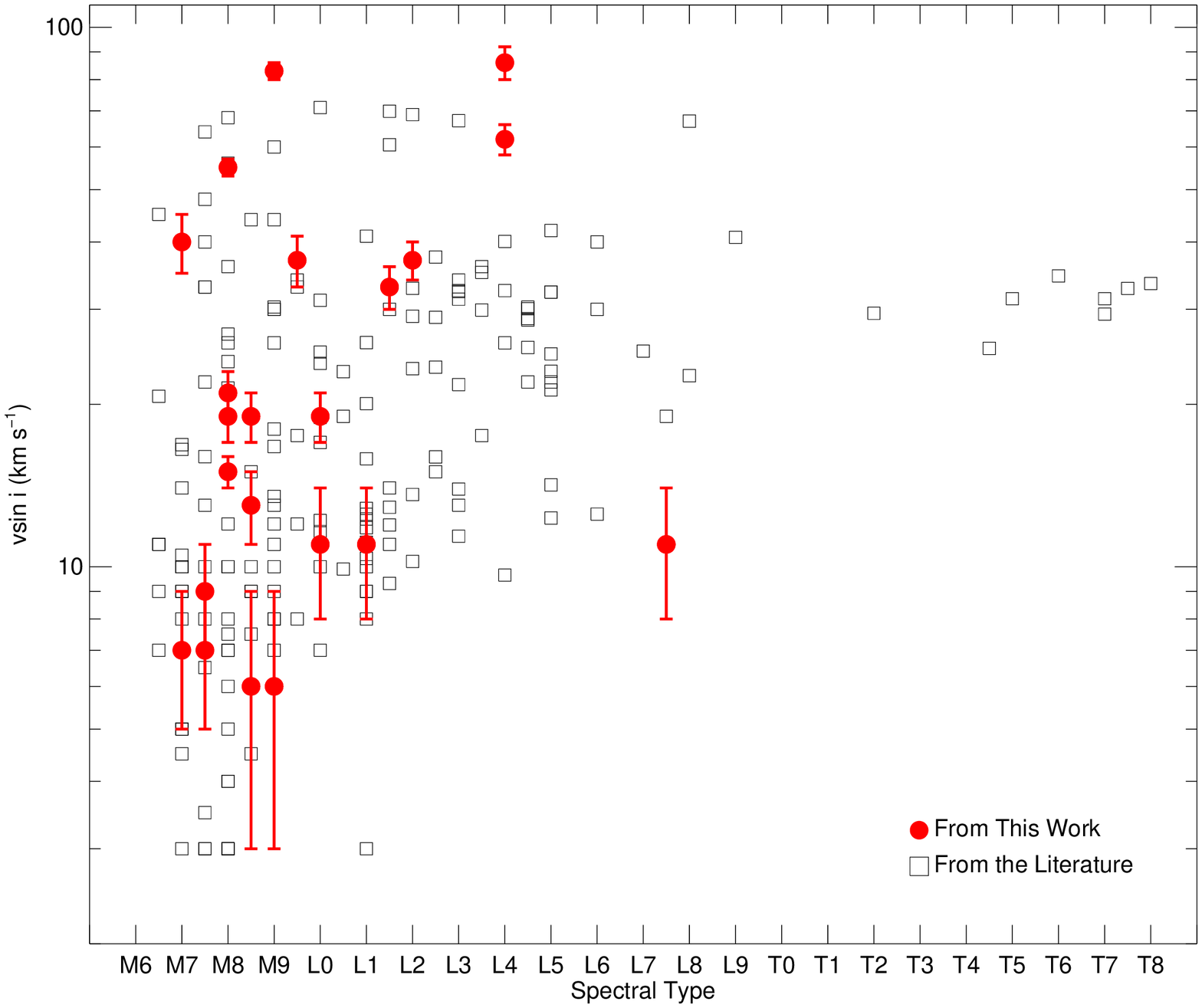,angle=0,width=\linewidth}
\caption{Projected rotational velocity
  ($v\,$sin$\,i$) versus spectral type for each of the binary
  components in our sample (red circles).  Also
  plotted (open squares, uncertainties omitted for clarity) are previously measured values (seeing limited
  observations, so binaries are not resolved) of $v\,$sin$\,i$ for
field VLM objects in the literature
\citep{mohanty03,zapo06,reiners08,reiners10,blake10}.  The values we measure 
for our sources are consistent with other VLM objects, with
our sources tending towards rapid rotation ($v\,$sin$\,i$ $>$10 km s$^{-1}$).}
\label{fig:st_all}
\end{figure}

In Figure \ref{fig:st_all}, we plot our measured $v\,$sin$\,i$s as
a function of spectral type.  We also include $v\,$sin$\,i$
measurements from the literature, derived from seeing-limited
observations \citep{mohanty03, zapo06, reiners08, reiners10, blake10} with
comparable sample, spectral resolution, and spectral type coverage to our
observations.  Our measurements are consistent with
previous observations, which also find that VLM objects tend
to be rapid rotators.  In addition, our results are
consistent with the trend of increasing $v\,$sin$\,i$ with
later spectral type.  In a number of cases, the values we
measure are higher than those objects presented in the literature of a
given spectral type.  This is likely attributable to the mixed ages
probed in this study.  \citet{reiners08} show that
rotational velocity is a function of the age of the system,
correlated with a spindown timescale that is driven by magnetic
braking.  In our sample, for instance, we find that the $v\,$sin$\,i$ measurements
for HD 130948BC are higher than all previous measurements for
mid-L dwarfs, and it has been proposed that this system is
younger than the majority of the field population
($\thicksim$400 - 800 Myr, \citealt{dupuy09a,mullan10}).
Indeed, the extremely rapid rotation of HD 130948BC may
imply that the younger age of the system preferred by
\citet{mullan10} is more plausible.  Our other extremely
rapidly rotating system, 
LP349-25AB, has also been proposed to be quite young
($\thicksim$140 Myr, \citealt{dupuy10}).  On the whole, however, the
measurements in our sample are fully consistent with the bulk
population of objects previously observed.

We also note that with these very rapid rotation rates,
both LP 349-25B and HD 130948C are rotating at
$\thicksim$30$\%$ of their break-up speed.  These large
rotational velocities 
should also cause a high degree of rotational flattening.  Using
equation 6 of \citet{barnes03}, we derive that
the ratio of the polar radii to the equatorial radii should
be $\thicksim$0.87.  Given this level of 
oblateness, these objects might 
be expected to exhibit measureable linear polarization
\citep{sengupta10}.  In addition, the 
cooler temperatures at the equator due to gravity darkening may affect 
the spectral type measurements, as
demonstrated by surface imaging of rapidly rotating
intermediate mass stars \citep{monnier07}.  The rapid rotation
also leads to a higher level of 
uncertainty in our radial velocity measurements for these
extremely rapidly rotating sources.  As is apparent in Figure
\ref{fig:spec_all}, the majority of the CO bandhead features
are basically smoothed by the rapid rotation, making anchoring
these objects precisely in wavelength space quite challenging
and giving rise to the relatively high radial velocity uncertainties given in
\citet{kono10}. 

Several sources in our sample were previously targeted in
studies that did not resolve the components, but did measure
$v\,$sin$\,i$.  LP 349-25AB, LHS 2397aAB, and 2MASS
2206-20AB were observed by \citet{reiners10}.  They obtained
values consistent with ours for 2MASS 2206-20AB and the
primary component of LP 349-25AB, but a very different value
for LHS 2397aAB.  We are not certain why our $v\,$sin$\,i$ measurement
of LHS 2397aAB is different from these authors, but speculate
that perhaps it can be attributed to the broadening of the
lines due to the binary orbit in the unresolved spectra.
\citet{jones96} measured the unresolved 
$v\,$sin$\,i$ for LP 415-20AB, also finding a result consistent 
with our measurements (for both components).  \citet{blake10}
measured an unresolved $v\,$sin$\,i$ for 2MASS 
0746+20AB.  They obtain a nearly identical value to what we measure
for 2MASS 0746+20B, although the combined 
light of the system should be dominated by the primary.
However, the flux ratio of $\thicksim$1.4 at K band means the 
dominance is not extreme.  

\subsection{Component $v\,$sin$\,i$ Comparison}

\begin{figure}
\epsscale{1.0}
\plotone{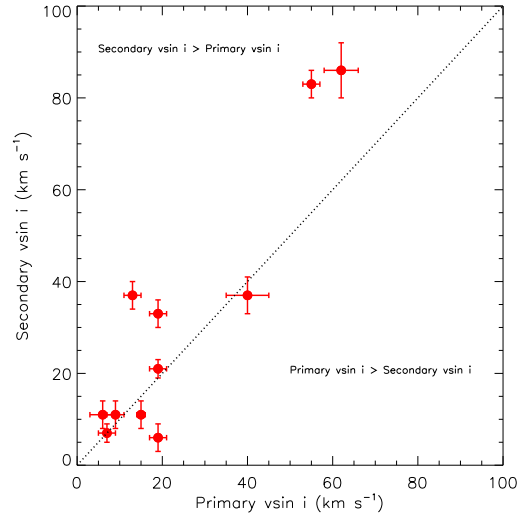}
\caption{The $v\,$sin$\,i$ of each secondary component
  plotted against its primary.  Sources with consistent
  velocities should fall on the dotted line.  Five of our eleven
  systems show components with $v\,$sin$\,i$ that differ by
  $>$3$\sigma$.  Of those five systems, four have secondary 
  components with higher velocities than their primaries.}
\label{fig:c12}
\end{figure}

We can also compare the rotational velocities of
the components in each system to each other.  The results of
this comparison are shown in Figure \ref{fig:c12}.  It is
immediately apparent that a number of components in the same
system have vastly different $v\,$sin$\,i$.  There are five binaries in our sample
that exhibit statistically-significantly ($>$3$\sigma$) differing $v\,$sin$\,i$.
These five systems, 2MASS 0746+20AB, 2MASS 2104+16AB, GJ
569Bab, HD 130948BC, and LP 349-25AB, have differences $>$10
km s$^{-1}$, or $>$30$\%$.  For all but GJ 569Bab, the secondary appears to
be rotating more rapidly 
than the primary.  For the other 6 systems in our sample, the
velocities are consistent to within 2$\sigma$.  We also note
that the consistency of these 6 systems, plus the generally
correlated rapid or slow rotation of the 5 systems with
different velocities, implies that the components of VLM
binaries are not randomly paired in $v\,$sin$\,i$.  

\begin{figure*}
\epsscale{1.0}
\plottwo{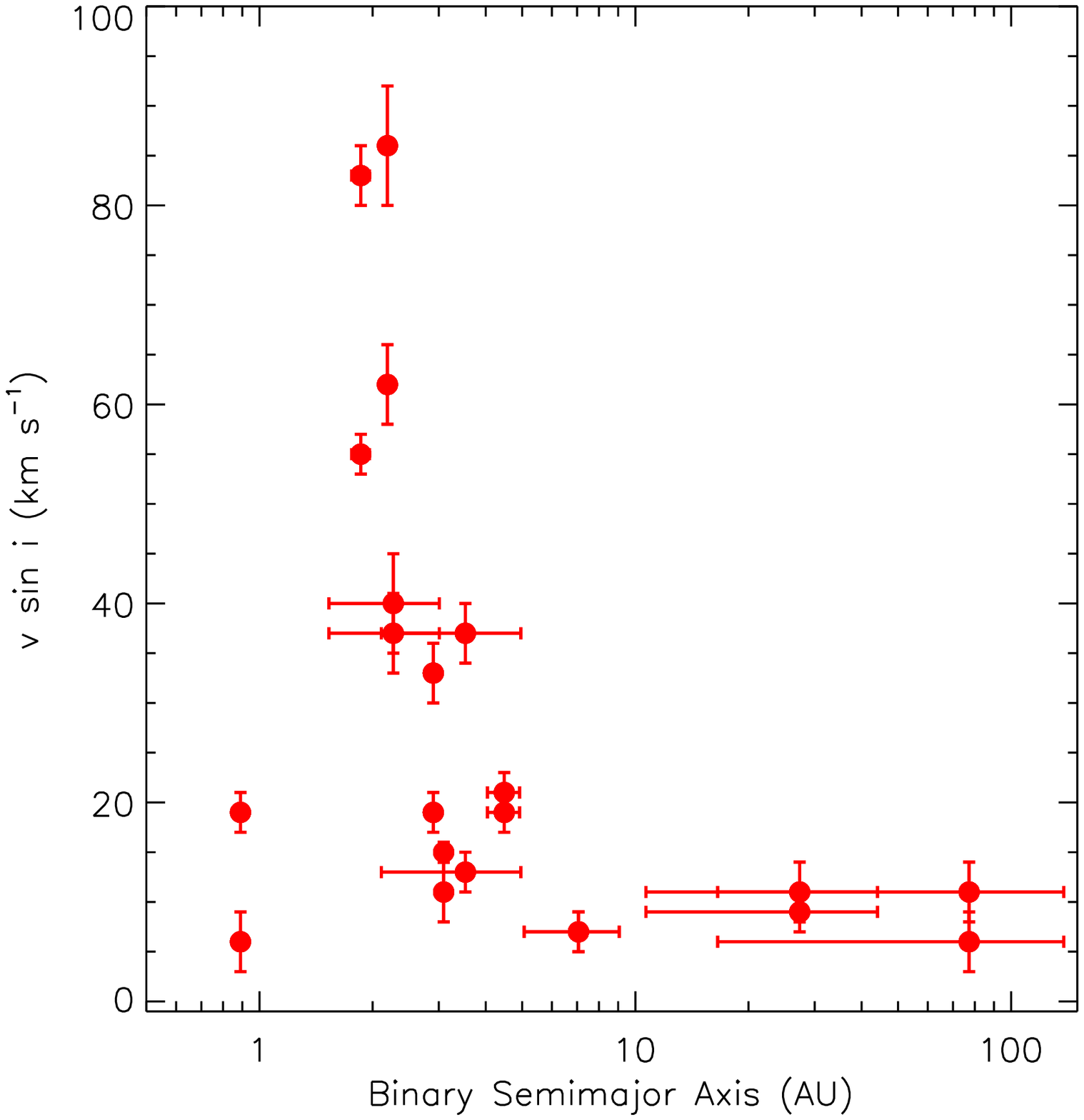}{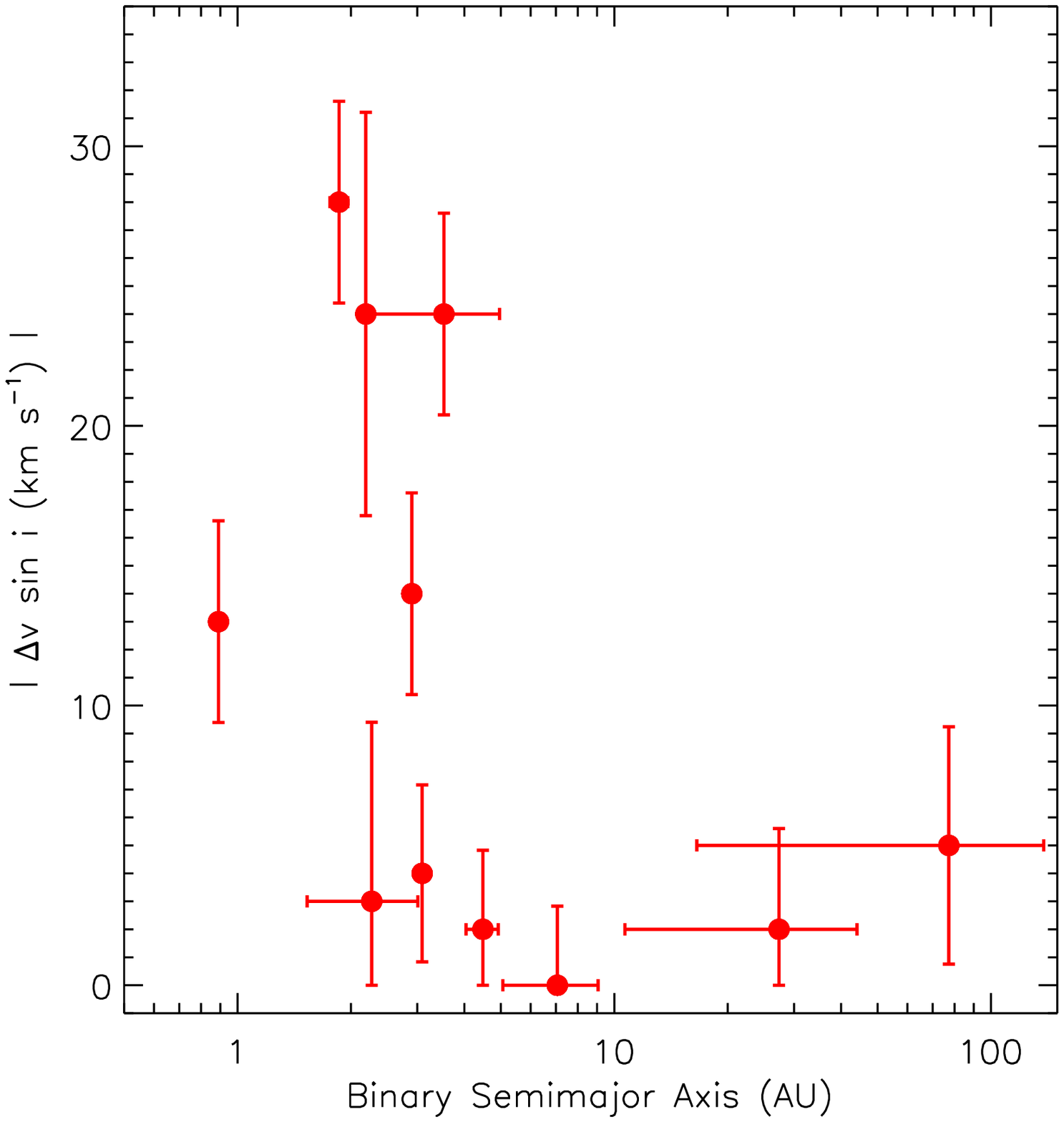}
\caption{The component $v\,$sin$\,i$ \textbf{(left)} and absolute value of the difference between
  component $v\,$sin$\,i$ \textbf{(right)} as a function of the binary semi-major
axis ($a$).  All systems with components having $v\,$sin$i\,$
$>$ 30 km s$^{-1}$ and with significant $v\,$sin$\,i$ differences ($>$3$\sigma$)
have $a$ $<$ 3.5 AU.  There also appears to be a rough trend of
decreasing $v\,$sin$\,i$ and $|\Delta$$v\,$sin$\,i|$ with increasing $a$.  The significance
of these trends given our current data and relatively small
sample is 2.3 - 2.9$\sigma$ for $v\,$sin$\,i$ vs. $a$ and 1.4
- 2.4$\sigma$ for $|\Delta$$v\,$sin$\,i|$ vs. $a$.}
\label{fig:sma_vsini}
\end{figure*}

\citet{simon06} noted that the lines of their
spatially-resolved spectra were broader for GJ 569Ba than for
GJ 569Bb.  They measured $v\,$sin$\,i$s of 25 km s$^{-1}$ and 10
km s$^{-1}$ for Ba and Bb, respectively, close to
the values we derive here.  The broadening of GJ 569Ba was
postulated to perhaps be due to an unresolved third
component rather than an intrinsic difference from GJ 569Bb.
Given that we see four other systems with differing component
$v\,$sin$\,i$, it is even more plausible that this system does not
have an unresolved third component.  In addition,
\citet{zapo04} measure $v\,$sin$\,i$s of 37 km s$^{-1}$ and 30
km s$^{-1}$ for Ba and Bb, thus also noticing a difference in
the component values.  The higher $v\,$sin$\,i$ values
potentially stems from their use of KI features in the J band
that are known to be gravity sensitive.  Although in many
cases the radial velocity uncertainties for these sources are
quite high \citep{kono10}, we do not see strong evidence for
additional radial velocity variability in any of these
discrepant systems that would point obviously to an additional unresolved
component.

We note that
in the case of HD 130948BC, \citet{mullan10} postulated that
the components may not be rotating at the same rate, which we
have now shown may be the case, although in contrast to
their predictions, the secondary is
likely the more rapid rotator and hence potentially the more
magnetically active component.  However, as we do not in fact
know the true equatorial velocity (v$_{eq}$) of either component, we
cannot make any definitive statements about 
their models, although the rapid rotation of both components
suggests that magnetic activity could be significant for both.
See Section \ref{sec:mutinc} for additional discussion of this
system.   

The targets in this sample all have orbital parameter estimates
from previous works
\citep{kono10,zapo04,simon06,dupuy09a,dupuy09b,dupuy09c,dupuy10}.  We therefore explore the potential impact of the binary orbital
properties on the consistency of $v\,$sin$\,i$.   For the purposes of this paper,
we use orbital elements derived in \citet{kono10}.   We looked for
trends in both $v\,$sin$\,i$ and 
$\Delta v\,$sin$\,i$ as a function of all orbital parameters, and the only variable that produces a
noticeable trend is semimajor axis ($a$).  All components with
$v\,$sin$\,i$ $>$ 30 km s$^{-1}$ and all five systems with
significantly different $v\,$sin$\,i$ have $a$ $\lesssim$ 3.5 AU.  In Figure
\ref{fig:sma_vsini}, we plot both the component $v\,$sin$\,i$
and $|\Delta v\,$sin$\,i|$
versus $a$.  To assess the significance
of the apparent trend of decreasing velocity and velocity difference with
increasing $a$, we use the Spearman's rank
correlation coefficient.  After accounting for the
uncertainties in $v\,$sin$\,i$ and $a$ via Monte
Carlo simulation, we determine using this metric that the
significance of the trend in $v\,$sin$\,i$ is between 2.3 and
2.9$\sigma$ and the significance in $|\Delta v\,$sin$\,i|$ is between 1.4 and 2.4$\sigma$.
Because of the relatively small size of our sample, we are
unable to explore the significance of this trend in greater
detail.  However, we note it here as a possible
relationship of interest in the context of the discussion
below.  In addition, this trend is similar to what was seen
in \citet{patience02}, who observed that tighter binaries in
$\alpha$ Per were rotating more rapidly than wider binaries.
However, such a trend was not observed by \citet{bouvier97} in
the Pleiades.

Given these result for the five ``discrepant'' systems, we are
confronted with two possible 
scenarios.  Either the binary components in some systems truly rotating at
significantly different rates, or their rotation axes are
inclined with respect to each other and possibly their orbital
plane (or some combination of these two).  We explore
these possibilities below in sections \ref{sec:intri} and \ref{sec:mutinc}.

\subsection{Intrinsic Rotational Velocity Differences?}\label{sec:intri}

\begin{figure}
%\epsscale{1.0}
\epsfig{file=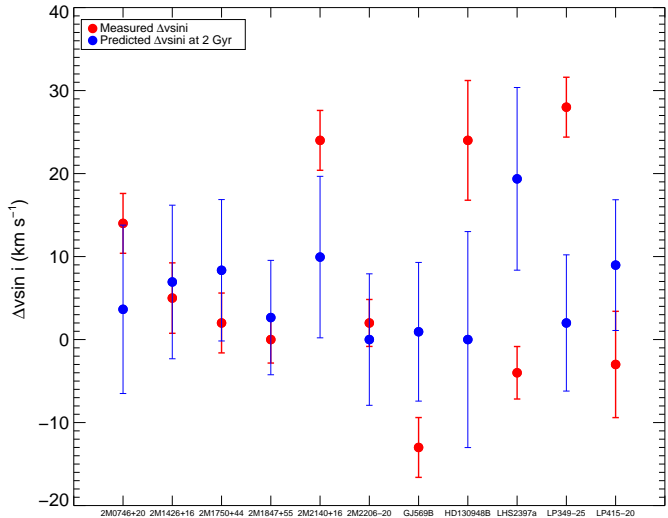,angle=90,width=\linewidth}
\caption{The measured and expected $\Delta$$v\,$sin$\,i$
  (secondary minus primary) for
  each binary based on the spectral type of the components.
  The expected value is derived using the relationship
  presented in \citet{reiners08} for $v\,$sin$\,i$ versus
  spectral type at an age of 2 Gyr.  We assume an uncertainty
  in derived $v\,$sin$\,i$ for each source of 20$\%$, which
  gives uncertainties in $\Delta$$v\,$sin$\,i$ of $\thicksim$7
  to 13 km s$^{-1}$.  While this relationship can
  account for the $\Delta$$v\,$sin$\,i$ of 2MASS 0746+20AB and
  2MASS 2140+16AB (marginally), the values of GJ 569Bab, HD 130948BC, and LP
  349-25AB are all $\gtrsim$1.6$\sigma$ discrepant.}
\label{fig:st_scatter}
\end{figure}

It is possible that the binaries in our sample with
``discrepant'' $v\,$sin$\,i$ have parallel rotation
axes but differing rotational velocities due to intrinsic
processes at work during either their formation or early evolution.
Binary systems are generally thought to form via fragmentation
of a molecular cloud core or large circumstellar disks wherein small seeds are formed that
eventually accrete more material, form a disk, and achieve
dynamical stability (e.g., \citealt{bonnell91, shu90}).
Although the formation mechanism for VLM objects is 
still an open question, a fragmentation origin is
certainly plausible for these objects.  Simulations of core 
fragmentation originally assumed that the rotation axes of the
binary seeds were aligned with the rotation axis of the core
\citep{bate97}.  In these simulations, it was shown that the
properties of the specific angular momentum of the accreting
material onto the protobinary had 
a substantial impact on the properties of the binary.  For a binary to grow to a mass
ratio of about 1, as in the case of most of the objects in our
sample, requires a higher specific angular momentum
for the accreting material, which in turn leads to a higher
accretion rate for the secondary component than the primary
\citep{bate97}.  The fraction of this angular momentum that is
converted into orbital angular momentum versus spin angular
momentum depends on the size of the accretion radius.  It has
been shown, however, that in cases where significant spin
increase is achieved from accretion, the primary component
tends to increase more than the secondary component
(\citealt{bate97,arty83}), which is 
not in agreement with most of our ``discrepant'' systems.  In 
addition, if these objects formed circumstellar disks after the
initial accretion phase, they may be subjected to braking
through magnetic coupling with the disks, as discussed below. 

As has been demonstrated in other works and discussed earlier,
there is a clear evolution of rotational velocity as a
function of age for VLM objects, much like what is observed
for higher mass stars \citep{reiners08}.  It has been
suggested that during the pre-main
sequence phase of evolution, rotation is likely
regulated by magnetic coupling to a circumstellar disk
(e.g., \citealt{edwards93, kundurthy06}), although
this mechanism is still a matter of debate (e.g.,
\citealt{stassun99, nguyen09}).  Once the disk dissipates in $\thicksim$1-10 Myr
\citep{strom93}, a star or VLM object will speed up due to its continued
contraction.  It has been shown that circumstellar disks exist
around components of binary stars (e.g., \citealt{mccabe06,
  cieza09}).  If component disks dissipate on different  
timescales, one object will begin to speed up sooner than the
other.  This
could potentially lead to differing rotational 
velocities in spite of coevality.  Observational support for
this possibility exists via measurements of disks around T
Tauri binaries.  For instance, \citet{mccabe06} identify six T Tauri
binary systems that have component disks in different phases
of evolution, with the secondary tending to have the more
evolved disk (albeit for slightly higher mass objects than
observed in this work).
However, it is unclear if a dissipation timescale difference,
which would likely be at most around 10 Myr, is alone sufficient to
generate a large rotational velocity difference in these systems.  
Furthermore, the tight physical separations of the binaries in our sample
should have impacted the formation and survival of any
circumstellar disks,
leading to both components having disks that last for at most
1 Myr (e.g., \citealt{cieza09, duchene10}).  The truncated disk
survival time in close binaries
could be related to the tentative trend we see in
$v\,$sin$\,i$ as a function of $a$ in Figure
\ref{fig:sma_vsini}.  

After the disk dissipates, VLMs likely follow a slightly modified form of the wind
braking law that has been shown to reproduce the rotation
versus age correlation seen amongst higher mass objects.  The
timescale for the braking of these objects is significantly
longer than for high mass objects.  In all cases, the rotational speed
seems to be clearly a function of mass, which in a coeval
system is correlated with spectral type.  It could be
postulated, therefore, that the reason we observe some
secondaries rotating more rapidly than primaries is due to
their lower mass/later spectral type.  Indeed, \citet{reiners07} identify this as
the cause of the $v\,$sin$\,i$ differences in the components of
the triple system LHS 1070.  In order to determine whether this
is a viable cause of the differences in $v\,$sin$\,i$ we observe
for our binaries, we attempt to roughly assess the expected
$v\,$sin$\,i$ as a function of spectral type for a given
age.  We use the relationship given in \citet{reiners08},
which uses the wind braking law to describe the 
rotational velocity evolution of VLM objects, for an age of 2
Gyr.  We choose this age because, although the ages of the
sources in our sample are generally unknown, it is roughly the average age
of the systems in our sample that are thought to be relatively
young ($\lesssim$500 Myr) and those that are thought to be the
age of the field (possibly as old as $\thicksim$5 Gyr).  
 Using this relationship, we
calculate the expected $v\,$sin$\,i$ for each binary component
based on its spectral type, and then compute the expected
difference between the two.  We also assume an intrinsic
scatter at a given spectral type to account for some of the spread
seen in measurements of rotation rates in clusters with
known ages (i.e., \citealt{terndrup00,irwin09}).  We assume a
conservative intrinsic scatter per 
object of 20$\%$, which gives an uncertainty
of between $\thicksim$7 and 13 km s$^{-1}$ in $\Delta$$v\,$sin$\,i$.
Because of the fairly substantial intrinsic scatter assumed
here, our choice of using the 2 Gyr relationship from
\citet{reiners08} has little impact on this comparison because
values of $\Delta$$v\,$sin$\,i$ from their 2, 5, and 10 Gyr
relationships are all consistent given these uncertainties
except in the case of LHS 2397aAB.  For this system, older
ages predict an even greater difference in
$\Delta$$v\,$sin$\,i$, whereas we observe no statistically
significant difference in $v\,$sin$\,i$ in this system.  The
only assumption that could potentially yield
$\Delta$$v\,$sin$\,i$ consistent with our measurement is a
very young age, which is most likely not the case for this
system (\citealt{freed03,dupuy09b}).  We plot
the values we derive for expected $\Delta$$v\,$sin$\,i$ in Figure
\ref{fig:st_scatter}, along with our actual measurements for
each binary.  While this effect, given our assumptions, could
explain the differences in the component velocities in 2MASS
0746+20AB and 2MASS 2140+16AB (marginally), it gives results inconsistent
with our measurements for GJ 569Bab (1.6$\sigma$ off), HD 130948BC
(1.6$\sigma$ off), or LP 349-25AB (2.9$\sigma$ off).  And, as
mentioned,
it predicts a large velocity difference for the components
of LHS 2397aAB, which we do not observe.  It is also worth
noting that GJ 569Bab, HD 130948BC, and LP 349-25AB have all
been postulated to be fairly young
($\lesssim$700 Myr, \citealt{dupuy09a,dupuy10,simon06,zapo06}).

\subsection{Mutually Inclined Rotation Axes?}\label{sec:mutinc}

\begin{figure}
\epsscale{1.2}
\plotone{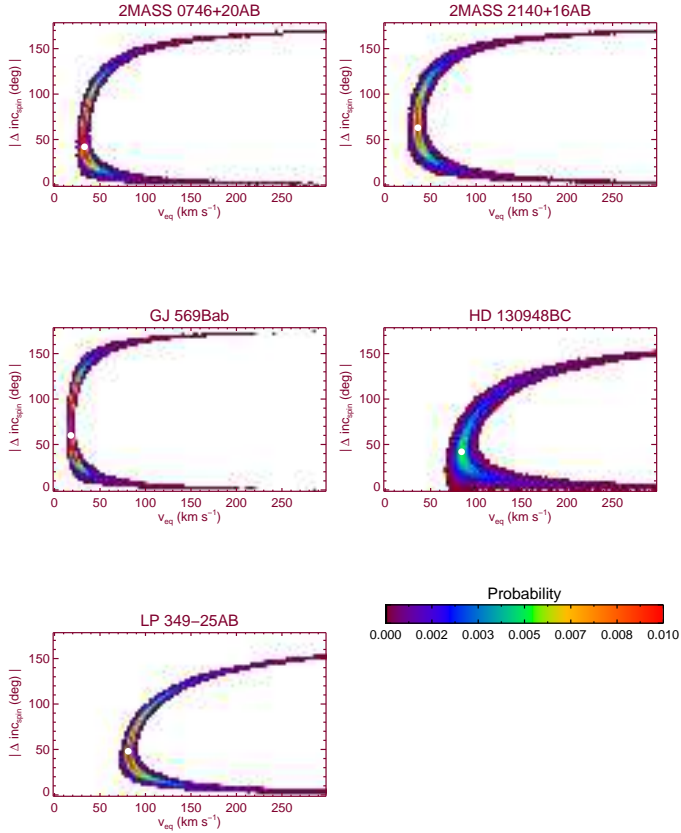}
\caption{Joint PDFs between the required relative inclination of the
  components of the five ``discrepant'' binaries and their
  assumed v$_{eq}$.  v$_{eq}$ is determined by 
  assigning the more rapidly rotating component an inclination
sampled from a distribution uniform in cos($i$) (ignoring the
known orbital inclination) and is assumed to be equal for both
components.  The white
circle on each PDF denotes the peak
probability.  All the peaks fall between 40 and 60$^{o}$.
Small values of $\Delta i$ generally require very
high v$_{eq}$.  The figures only extend to 300 km s$^{-1}$,
which is roughly the break up speed for these objects.}
\label{fig:i12}
\end{figure}

If we instead assume that the binary components with similar masses must have
similar or the same rotational velocities, then their rotation
axes must be inclined with respect to one another in systems
with differing $v\,$sin$\,i$.  To determine the most likely
value of the relative inclination of the spin axes in these
five systems, we ignore
for a moment the known orbital inclination and perform a Monte
Carlo simulation assuming one component has an inclination
sampled from a distribution that is uniform in cos($i$).  We
select a value of $v\,$sin$\,i$ for each component from a
Gaussian distribution defined by our measurements and
uncertainties, and assign the randomly sampled inclination to
the faster rotating component to determine the equatorial
velocity (v$_{eq}$) for the
system.  We then calculate the inclination required for the
other component to have the same v$_{eq}$.  We perform this
exercise 100000 times to derive probabilities, which we plot
in Figure \ref{fig:i12}.  Note that there are two possible
values for the component with ``unknown'' inclination that
will give v$_{eq}$ due to a 180$^{o}$ ambiguity (axis pointing
up or down from our line of sight).  Figure \ref{fig:i12}
shows joint probability density functions (PDFs) between v$_{eq}$ and the
absolute value of the difference in the inclination of the two
components.  The white dot overplotted on each distribution
denotes the location of peak probability.   These peaks are
found between $\thicksim$40 and $\thicksim$60 degrees in
$\Delta i$ and fall close to a v$_{eq}$ that equals
the $v\,$sin$\,i$ of the
faster rotating component since a randomly oriented object is
more likely to be observed edge-on than pole-on.  The figure
also demonstrates that the only configuration that can
maintain both the same inclination and v$_{eq}$ is one in
which both components rotate close to break up speed
($\thicksim$300 km s$^{-1}$) \textbf{and} are observed almost
pole on.  This is an extremely unlikely configuration and can
safely be ruled out as an explanation for all the systems with
discrepant $v\,$sin$\,i$.

It is possible that some level of rotation axis misalignment
is natural in these binary systems and could be represented by
a Gaussian distribution centered on some ``typical''
misalignment.  We can use our sample to assess this toy
model assuming the components have the same v$_{eq}$.  In order to determine the 
value for the typical relative inclination and its 1$\sigma$
spread, we performed another Monte Carlo simulation in which we
sampled a relative inclination for each system from the
distribution allowed by our $v\,$sin$\,i$ measurements and
uncertainties.  In this case, however, we assumed that one
component was aligned with the orbital plane, sampling from a
distribution allowed by the measurements of orbital
inclination \citep{kono10}.  These
inclination distributions tend to avoid cases with very high
v$_{eq}$ and therefore are more realistic.
As before, we assigned the more rapidly
rotating component the sampled inclination, and calculated the
inclination required for the 
other component to have the same v$_{eq}$.  We then
generated many (100000) distributions of relative inclinations
for our sample, which we fit with a simple Gaussian model to
find the peak and full width half maximum.  We performed the
simulation for two cases, one in which we chose the smaller of
the two allowed relative inclinations (due to the 180$^{0}$
ambiguity), and the other chosing at random either the smaller
or the larger of the the allowed relative inclinations.  In
the first case (small angles) we find that the preferred
distribution has the form 18 $\pm$ 26$^{0}$ and in the
second case (any angle) the distribution has the form 25
$\pm$ 60$^{0}$.  The implication of this is that if we
are truly probing a Gaussian distribution of relative
inclinations with the expectation of a few many-sigma
outliers, the average inclination and spread must be quite
substantial in order to describe for our sample.  
 
The idea of forming objects in which the rotation axis is 
inclined with respect to the orbital axis has been explored in
great detail recently due to the discovery of planets orbiting
in a plane that is misaligned with the stellar rotation axis
(e.g., \citealt{winn09, winn10, triaud10}).  For instance,
\citet{bate10} investigated 
the accretion history of stars forming in a turbulent cluster
environment.  Given variable accretion rates and material
being accreted from different directions, it is possible to
impact both the rate of rotation and the axis of rotation,
leading to rotation axes that are misaligned with
circumstellar disks.   Alternatively,
\citet{lai11} have explored how the interaction between the
magnetic field of a young star and its circumstellar disk
can effectively push the stellar rotation axis out of
alignment due to a magnetic warping torque.  Such scenarios
could potentially cause binary stars to have inclined rotation
axes.

\begin{figure}
\epsscale{1.0}
\plotone{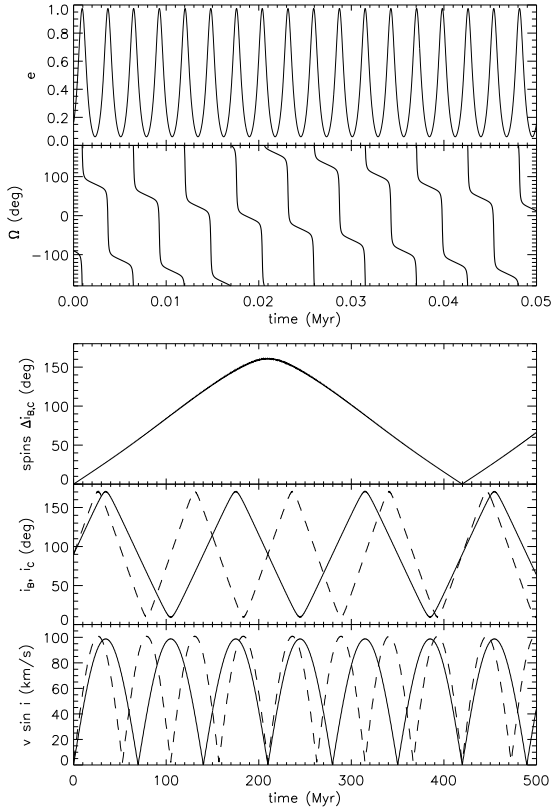}
\caption{\textbf{Top two panels:}
Eccentricity oscillations and regression of the ascending node
(in the reference frame of the A-BC orbit) for HD 130948BC due
to the presence of HD 130948A.  Based on the observed and
modeled properties of the system, it could be undergoing
\citet{kozai62} oscillations on a $\sim10^4$~yr period.   
\textbf{Bottom three panels:}
Evolution of the angular difference in the spin vectors
($\Delta i_{B,C}$), rotation axis inclination to the line of
sight, and a possible resulting time history of $v\,$sin$\,i$ of
HD 130948B (solid line) and  C (dashed line) due to the
presence of HD 130948A.  This effect could explain our
$v\,$sin$\,i$ measurements and do so within the expected lifetime of the system
($\thicksim$500 Myr.)} 
\label{fig:precess}
\end{figure}

A third possibility is secular torques due to a third stellar
companion.  Indeed, \citet{hale94} gave a general rule that
binary components closer than 40 AU have agreement in spin
orientation, but mentioned that rule is broken for systems
with a third body.  In particular, consider a binary star that
is born with each of its component spin vectors aligned with
the orbital angular momentum.  If the orbital orientation
remains fixed throughout the binary's lifetime, there would be
no torque to cause spin precession, and the components would
remain aligned with the orbit.  However, if a companion star
($m_3$) is introduced on an external, non-coplanar orbit of
period $P_{\rm out}$ and inclination $\,i$ relative to the
binary ($m_1$-$m_2$) of period $P_{\rm in}$, it will cause
orbital nodal precession in the binary on a secular period:  

\begin{equation}
T_{\rm sec} \approx \frac{4}{3 \cos i} \frac{P_{\rm out}^2} {P_{\rm in}} \frac{m_1+m_2+m_3}{m_3}.  \label{eqn:psec}
\end{equation}
\citep{kiseleva98}.  In response to this misaligned orbit, the spins will precess as well, on a period:

\begin{equation}
T_{\rm precess,1} = P_1 \frac{m_1}{m_2} \frac{C_1}{k_{\rm 2,1}} \bigg( \frac{a}{R_1} \bigg)^3  \frac{ (1-e^2)^{3/2} }{ \cos \psi }  \label{eqn:precess}
\end{equation}
\citep{egg01} for star 1, where $k_{2,1}$ is the apsidal motion constant
$k_2$ for star 1, $C_1$ is its normalized inertia ($=I/MR^2$),
$P_1$ is its spin period, $\psi$ is its spin obliquity, and
$a$ and $e$ are the semi-major axis and the eccentricity of
the orbit.  An analogous period for $m_2$ need not be the
same if the masses and radii are not quite equal.  Thus we
have the opportunity to begin a binary aligned, and due to
slight differences in precession rate, open up a large
inclination difference between the component spin axes.
Thus, for the spins to become misaligned with one another, we
require 

\begin{equation}
\left| \frac{1}{T_{\rm precess,1}} - \frac{1}{T_{\rm precess,2}} \right| = \frac{1}{\rm system\, age }.
\end{equation}

However, if the two spin precession periods are different
but much shorter than the orbit precession period
(equation~\ref{eqn:psec}), then the spins will just track the
orbit orientation and will not become misaligned from either
the orbit or each other.  Therefore, we also require: 

\begin{equation}
T_{\rm sec} \lesssim max(T_{\rm  precess,1},T_{\rm precess,2}).
\end{equation}

In our sample, we have two binaries with similar spectral
types (GJ 569Bab and HD
130948BC) that are part of known triple systems.  We have
therefore produced a numerical demonstration of this effect
for HD 130948BC, which has component $v\,$sin$\,i$ values that
differ, despite similar luminosities and spectral type.
Although the mass ratio of this system has not been directly
measured, we can follow the method of \citet{dupuy09a} and use
a system age estimate and the component bolometric
luminosities to estimate individual masses.  
This method predicts a slight mass difference due to the factor of
$\thicksim$1.1 difference in the luminosities.  
While the components should have very similar radii now, this
was not the case when they were younger if they do indeed have
slightly different masses.  Because the radius is such an
important factor in equation \ref{eqn:precess}, we assume for
the purposes of this demonstration the average
radii for these objects over the approximate age of the system
($\thicksim$500 Myr, \citet{dupuy09a}): $(R_1, R_2)=(0.158,
0.150)R_\odot$ (\citealt{chabrier00}, DUSTY models), and corresponding masses $(M_1, M_2) = (0.06,
0.05) M_\odot$ (\citealt{kono10, dupuy09a}).  For the sake of this argument, we set 
$P_1=1.92$ hours and $P_2=1.82$ hours (i.e., v$_{eq}$ = 100 km s$^{-1}$), so that the observed difference in
$v\,$sin$\,i$ will be due to inclination differences rather than
rotation rate differences.  Last, we chose apsidal motion
constants $k_{2,1}=k_{2,2}=0.175$ and normalized moment of
inertia $C_1=C_2=0.108$ \citep{leconte11}.  We integrated the binary orbital
parameters (starting with the observed $a=2.19$~AU and
$e=0.16$, \citealt{kono10}) in the gravitational potential of the third star
($m_3 = 1.1 M_\odot$, $a=50$~AU, $e=0.7$ in an orientation
$i=80^\circ$, $\omega_{\rm in}=45^\circ$), using a secular code
that uses the quadrupole approximation for the dynamics of the
three stars (\citealt{fab07, egg01}).  In Figure
\ref{fig:precess} we show the results.  On the short period
$T_{\rm sec} \approx 10^4$~yr, the inner orbit precesses
and goes through eccentricity oscillations, i.e.,
\citet{kozai62} cycles.  On the longer period $T_{\rm precess} \approx 4\times10^8$~yr, the spins precess, and on
an even longer period, $1/(1/T_{\rm precess,1}-1/\tau_{\rm precess,2}) \approx1\times10^9$~yr,
the precession rates mix in phase (one gains by $\pi$, due to
the slightly different masses and radii). The system is likely
as young as 500 - 700 Myr 
(\citealt{dupuy09a,mullan10}).  Therefore it is plausible that
the differential precession only recently opened a measurable
angle between the rotation axes of these two components.  In any case,
this exercise suggests that no matter the true masses and
radii of the components, Kozai cycles are quite possibly
occurring in this system, and the differences in component
$v\,$sin$\,i$s are possibly just one manifestation of the
precession of the inner binary.  The rapidity of the
oscillation period implies that the precession of the inner
orbit and/or its eccentricity oscillation may be measurable;
e.g., the eccentricity currently has a precision of
$\thicksim$0.01 (\citealt{dupuy09a,kono10}) and this numerical
integration shows eccentricity changes of 0.01 per decade. 

We have also performed an analogous simulation for the GJ 569Bab
system, which exhibits $v\,$sin$\,i$ differences of only 13 km
s$^{-1}$ as opposed to the 24 km s$^{-1}$ difference in HD
130948BC.  Using the orbital parameters of GJ 569Bab, which
has a tighter separation ($\thicksim$1 AU), we find that the precession
period for the system is much more rapid, on the order of
0.1 Myr.  It is therefore substantially easier to have
precession occur for this system, making it 
plausible that secular perturbations are responsible for the
$v\,$sin$\,i$ difference in this system as well. 

Three of the five systems with differing $v\,$sin$\,i$s
are not part of known triple systems.  \citet{allen07}
performed deep imaging around 2MASS0746+20AB and
2MASS2140+16AB and found no comoving companions between
40-1000 AU down to a mass limit of $\thicksim$0.05
M$_{\odot}$.  Although we are unaware of similar deep imaging
for LP 349-25AB, current all-sky surveys do not reveal any
bright sources within 1000 AU.  This does not,
however, rule out the possibility that these sources were
previously members of higher order multiple systems, or had an
interaction with an unrelated object.  We also note that to
our knowledge, none of the sources in our sample with
consistent $v\,$sin$\,i$ have additional companions.

\subsection{Implications for Radio Observations}

Two of our targets (2MASS0746+20AB and LP 349-25AB) are
known radio sources \citep{antonova07,phan07}.  The components
of these binaries 
exhibit rapid rotation ($\gtrsim$19 km s$^{-1}$).   Both systems also have components with
different $v\,$sin$\,i$.  This has interesting implications for
determining which of the binary components is emitting in the
radio, as the radio measurements are typically
unresolved.  For instance, \citet{berger09} used a
$v\,$sin$\,i$ for 2MASS 0746+20AB from unresolved
measurements to derive a radius for the radio emitting
component, which they assumed to be the primary.  Using the spatially
resolved measurements of $v\,$sin$\,i$ for these objects rather than
an unresolved value (which at 27 km s$^{-1}$ is nearly
exactly the average of 
the spatially resolved values of 19 and 33 km s$^{-1}$ that we
obtain here) gives slightly different values for the
radius.  We can now derive the predicted radius for each
component without assuming which is the emitting source and
determine which gives a more plausible result.  If we make the same assumption as \citet{berger09} that
the rotation axis is perpendicular to the orbital plane, which was
updated by \citet{kono10} to have an inclination of 138.2 $\pm$
0.5$^{o}$, we derive velocities of 29 $\pm$ 3 km s$^{-1}$ and 50 $\pm$
3 km s$^{-1}$ for the primary and secondary, respectively.  Using the rotational period from
\citet{berger09} of 124.32 $\pm$ 0.11 minutes gives a radius
of 0.050 $\pm$ 0.005 R$_{\sun}$ for the primary and 0.085 $\pm$ 0.005 R$_{\sun}$ for the
secondary.  Assuming the total system mass and
preliminary estimates of the component mass from
\citet{kono10} that the mass of the primary is between 0.08
and 0.1 M$_{\sun}$ while the secondary is between 0.05
and 0.07 M$_{\sun}$, evolutionary models predict radii of
0.103 - 0.125 R$_{\sun}$ for for the primary and 0.094 - 0.096
R$_{\sun}$ for the secondary.  The implied discrepancy with
the models is therefore 10$\sigma$ for the primary and  
1.8$\sigma$ for the secondary.  Given this, in conjunction
with the unphysically small radius predicted for the primary, we postulate that
the secondary is the source of the radio emission, which may
be related to its more rapid rotation.  Although 
the radius of 0.085 $\pm$ 0.005 R$_{\sun}$  
is consistent to within the uncertainties of the value from
\citet{berger09}, the resulting overprediction of the radius
by evolutionary models becomes slightly less severe because if
the emission is from the secondary, the models predict it will
have a smaller
radius.  In either case, this analysis highlights the importance 
of obtaining fundamental parameters of binary components
individually rather than from unresolved measurements.  In
addition, the results from our sample overall suggest that the assumption of
rotational axis alignment with the orbital plane should be
treated with much greater caution.   

We can also speculate that the secondary
component of LP 349-25AB is likely the radio source in the system, since it is
rotating extremely rapidly.  However, it is not improbable
that both components are radio sources.  Future VLBI
observations will probe the true origin of the radio emission
in both of these system (G. Hallinan et al., in prep).
Furthermore, our results provide a guideline by which sources
previously unobserved in the radio should be targeted.
Binaries that have at least one component with $v\,$sin$\,i$ $>$
30 km s$^{-1}$ would make excellent candidates for observation.

\section{Summary}\label{conc}

Using the combination of high spatial and high spectral
resolution afforded by the Keck II LGS AO system, we have
measured component $v\,$sin$\,i$s for a sample 
of 11 VLM binaries.  Among the 22 objects measured, 80$\%$ are
rapid rotators ($v\,$sin$\,i$ $\gtrsim$10 km s$^{-1}$), consistent with previous measurements for VLM
objects.  We found that 5 of the binaries surveyed
had components with $v\,$sin$\,i$'s that differed by $>$3$\sigma$.  We explored potential causes for these differences,
which must stem either from intrinsic velocity differences or
from mutually inclined rotation axes, or a combination of both.  Our analysis shows that perhaps both explanations are required
to explain these five binaries, with two (2MASS 0746+20AB and
2MASS 2140+16AB) having explainable
intrinsic differences due to differing spectral types, while the
other three (GJ569Bab, HD 130948BC, and LP 349-25AB) likely require a different explanation.  We
looked at the binary HD 130948BC as an example of a
case in which secular torques are most likely causing the
orbit and component spin vectors to evolve.  This binary,
in which Kozai oscillations are possibly at work,
is an example of the impact of dynamical evolution on
VLM binaries.  One other binary in our sample with differing
component velocities, GJ 569Bab, is part of a triple system
and a similar analysis shows it too may be displaying the
impact of secular torques.
For LP 349-25AB, which is not part of a known
triple system and spectral type differences do not seem to
account for its vastly different $v\,$sin$\,i$s, the explanation may rest with past dynamical
encounters or perturbations by higher mass objects.  
Our results also have implications for the previous
measurements of radio activity in this system and 2MASS
0746+20AB.  We suggest that the secondary components are more likely
the radio emitting sources in these systems.  This stresses
the importance of measuring fundamental parameters of binary
components individually rather than bootstrapping from
unresolved measurements.

Continued monitoring of these systems will improve the
precision to which parameters such as component mass and
radius are measured, allowing for the correlation of the
properties with rotational velocity.  This will provide a new
handle on the ages of these objects.  Further, the objects
that we found to be rapid rotators that have not yet been
surveyed for radio emission are ideal targets for observation. 

\acknowledgements

The authors thank observing assistants Joel Aycock, Heather
Hershley, Carolyn Parker, Gary Puniwai, Chuck Sorenson,
Terry Stickel, and Cynthia Wilburn and support astronomers Randy
Campbell, Al Conrad, Marc Kassis, Jim Lyke, and Hien Tran for
their help in obtaining the observations.  We thank John
Bailey for helpful advice regarding the analysis.  We also thank an
anonymous referee for helpful suggestions for the improvement
of this document.  This work was
performed under the auspices of the U.S. Department of Energy
by Lawrence Livermore National Laboratory under Contract
DE-AC52-07NA27344.  Support for this 
work was provided by the NASA Origins Program (NNX1 OAH39G) and the NSF Science
\& Technology Center for AO, managed by UCSC
(AST-9876783).  Some of the spectral analysis tools used in
this study were developed in part by funding provided by
NSF/AAG Grant \#0908018.  This
publication makes use of data products from the Two Micron All
Sky Survey, which is a joint project of the University of
Massachusetts and the Infrared Processing and Analysis
Center/California Institute of Technology, funded by the
National Aeronautics and Space Administration and the National
Science Foundation.  The W.M. Keck Observatory is operated as
a scientific partnership among the California Institute of
Technology, the University of California and the National
Aeronautics and Space Administration. The Observatory was made
possible by the generous financial support of the W.M. Keck
Foundation.  The authors also wish to recognize and
acknowledge the very significant cultural role and reverence
that the summit of Mauna Kea has always had within the
indigenous Hawaiian community.  We are most fortunate to have
the opportunity to conduct observations from this mountain.

\end{document}

%% file: tab1.tex
%\begin{singlespace} 
\begin{deluxetable*}{lccccc} 
\tabletypesize{\scriptsize} 
\tablewidth{0pt} 
%\rotate
\tablecaption{VLM Binary Sample} 
\tablehead{ 
  \colhead{Source Name} & \colhead{RA} & \colhead{Dec} &
  \colhead{Estimated} & \colhead{Discovery} &
  \colhead{2MASS}\\
  \colhead{} & \colhead{(J2000)} & \colhead{(J2000)} &
  \colhead{Sp Types\tablenotemark{a}} & \colhead{Reference} & \colhead{K Band Mag.}
}
\startdata 
LP 349-25AB & 00 27 55.93 & +22 19 32.8 & M8+M9 & 1 & 9.569 $\pm$ 0.017\\
LP 415-20AB & 04 21 49.0 & +19 29 10 & M7+M9.5 & 2 & 11.668 $\pm$ 0.020\\
2MASS J07464256+2000321AB & 07 46 42.5 & +20 00 32 & L0+L1.5 & 3 & 10.468 $\pm$ 0.022\\
GJ 569Bab  & 14 54 29.0 & +16 06 05 & M8.5+M9 & 4 &$\sim$9.8 \\
LHS 2397aAB & 11 21 49.25 & -13 13 08.4 & M8+L7.5 & 5 & 10.735 $\pm$ 0.023 \\
2MASS J14263161+1557012AB & 14 26 31.62 & +15 57 01.3 & M8.5+L1 & 6 & 11.731 $\pm$ 0.018 \\
HD 130948BC & 14 50 15.81 & +23 54 42.6 & L4+L4 & 7 & $\sim$11.0\\
2MASS J17501291+4424043AB & 17 50 12.91 & +44 24 04.3 & M7.5+L0 & 2 & 11.768 $\pm$ 0.017\\
2MASS J18470342+5522433AB & 18 47 03.42 & +55 22 43.3 & M7+M7.5 & 8 & 10.901 $\pm$ 0.020 \\
2MASS J21402931+1625183AB & 21 40 29.32 & +16 25 18.3 & M8.5+L2 & 6 & 11.826 $\pm$ 0.031 \\
2MASS J22062280-2047058AB & 22 06 22.80 & -20 47 05.9 & M8+M8 & 6 & 11.315 $\pm$ 0.027\\
\enddata
\tablenotetext{a}{From discovery reference}
\tablecomments{References - (1) Forveille et al. 2005 (2)
Siegler et al. 2003 (3) Reid et al. 2001 (4) Mart{\'{\i}}n et
al. 2000 (5) Freed et al. 2003 (6) Close et al. 2003 (7) Potter et
al. (2002) (8) Siegler et al. 2005}
\label{tab:sample}
\end{deluxetable*} 
%\end{singlespace} 

%% file: tab2.tex
%\begin{singlespace}
\begin{deluxetable*}{lcccccc}
\tabletypesize{\scriptsize}
\tablecolumns{7}
\tablewidth{0pc}
\tablecaption{Log of NIRSPAO LGS K-band Observations\tablenotemark{a}}
\tablehead{
\colhead{Target Name} &  \colhead{Date of} & \colhead{A0V Star} &
\colhead{Exposure Time} & \colhead{No. of} &
\colhead{Avg. SNR} & \colhead{Avg. SNR} \\
\colhead{} &\colhead{Observation (UT)} &  \colhead{Standard} & \colhead{(sec)} &
\colhead{Frames} & \colhead{Primary} & \colhead{Secondary}
}
\startdata
2MASS J07464256+2000321AB & 2006 Dec 16 & HIP 41798 & 1200 & 4 & 52 & 44\\
                          & 2007 Dec 04 & HIP 41798 & 1200 &6 & 72 & 59\\
                          & 2008 Dec 19 & HIP 41798 & 1200 & 6 & 66 & 56\\
                          & 2009 Dec 09 & HIP 41798 & 1200 & 2 & 83 & 70\\
2MASS J14263161+1557012AB & 2007 Jun 08 & HIP 73087 & 1200 & 4 & 44 & 33\\ 
                          & 2008 Jun 01 & HIP 73087 & 1200 & 4 & 50 & 36 \\
                          & 2009 Jun 12 & HIP 73087 & 1200 & 4 & 41 & 29 \\
2MASS J17501291+4424043AB & 2008 May 31 & HIP 87045 & 1200 & 4 & 48 & 36 \\
                          & 2009 Jun 12 & HIP 87045 & 1200 & 6 & 41 & 31 \\
                          & 2010 Jun 07 & HIP 87045 & 1200 & 4 & 36 & 26 \\
2MASS J18470342+5522433AB & 2007 Jun 08 & HIP 93713 & 1200 & 4 & 69 & 60\\
                          & 2008 Jun 01 & HIP 93713 & 1200 & 5 & 69 & 60 \\
                          & 2009 Jun 13 & HIP 93713 & 1200 & 3 & 39 & 36 \\
                          & 2010 Jun 07 & HIP 93713 & 1200 & 4 & 53 & 47 \\
2MASS J21402931+1625183AB & 2007 Jun 09 & HIP 108060 & 1200 & 4 & 43 & 28 \\
                          & 2008 May 31 & HIP 108060 & 1800 & 3 & 58 & 40\\
                          & 2009 Jun 13 & HIP 108060 & 1800 & 2 & 38 & 26 \\
                          & 2009 Dec 10 & HIP 116611 & 1800 & 3& 56 & 45\\
2MASS J22062280-2047058AB & 2007 Jun 09 & HIP 116750 & 1200 & 3 & 47 & 39\\
                          & 2008 Jun 01 & HIP 109689 & 1200 & 4 & 54 & 48 \\
                          & 2009 Jun 12 & HIP 109689 & 1200 & 4 & 47 & 44 \\
                          & 2010 Jun 07 & HIP 109689 & 1200 & 4 & 29 & 29 \\
GJ 569Bab                 & 2007 Jun 09 & HIP 73087 & 900 & 2 & 89 & 82 \\
                          & 2009 Jun 13 & HIP 73087 & 900 & 4 & 86 & 67 \\
                          & 2010 Jun 06 & HIP 73087 & 900 & 4& 114 & 97 \\
HD 130948BC               & 2007 Jun 09 & HIP 73087 & 1200 & 4 & 43 & 37\\
                          & 2010 Jun 07 & HIP 73087 & 1800 & 4 & 59 & 52 \\
                          & 2011 Jun 18 & HIP 73087 & 1800 & 4 & 52 & 48 \\
LHS 2397aAB               & 2007 Dec 04 & HIP 58188 & 1800 & 2 & 68 & 27\\
                          & 2008 May 31 & HIP 61318 & 1800 & 3 & 114 & 44\\
                          & 2008 Dec 19 & HIP 58188 & 1800 & 3 & 83 & 31 \\
                          & 2009 Jun 12 & HIP 61318 & 1800 & 2 & 103 & 33 \\
                          & 2009 Dec 09 & HIP 58188 & 1800 & 3 & 105 & 29 \\
                          & 2010 Jun 07 & HIP 61318 & 1800 & 2 & 77 & 31 \\
LP 349-25AB               & 2006 Dec 16 & HIP 5132 & 600 & 4 & 58& 45\\
                          & 2007 Dec 04 & HIP 5132 & 900 &1 & 63 & 58\\
                          & 2008 Dec 19 & HIP 5132 & 1200 & 4 & 105 & 84\\
                          & 2009 Jun 12 & HIP 5132 & 1200 & 4 & 114 & 98 \\
                          & 2009 Dec 09 & HIP 5132 & 1200 & 4 & 121 & 107 \\
LP 415-20AB               & 2008 Dec 19 & HIP 24555 & 1200 & 4 & 42 & 32\\
                          & 2009 Dec 09 & HIP 22845 & 1800 & 2& 62 & 49 \\
\enddata
\tablenotetext{a}{All data taken before December 2009
  represents the same NIRSPAO-LGS data set presented in \citet{kono10}}
\label{tab:obslog}
\end{deluxetable*}
%\end{singlespace}

%% file: tab3.tex
%\begin{singlespace}
%\begin{landscape}
\tabletypesize{\scriptsize}
\begin{deluxetable*}{lccccccccccc|c}
%\tabletypesize{\scriptsize}
\setlength{\tabcolsep}{1.0mm}
%\rotate
\tablecolumns{9}
\tablewidth{0pc}
\tablecaption{$v\,$sin$\,i$ Measurements (km s$^{-1}$)}
\tablehead{
\colhead{Target} & \colhead{Sp.} & \colhead{Adopted} &\colhead{2006} & \colhead{2007}
& \colhead{2007} & \colhead{2008} &
\colhead{2008} & \colhead{2009} & \colhead{2009}
& \colhead{2010}& \colhead{2011}\vline & \colhead{Weighted} \\
\colhead{} & \colhead{Type} & \colhead{T$_{eff}$ (K)\tablenotemark{a}} &\colhead{December} & \colhead{June}
& \colhead{December} & \colhead{May/June} &
\colhead{December} & \colhead{June} & \colhead{December}
& \colhead{June} &\colhead{June}\vline & \colhead{Average} 
}
\startdata
2MASS 0746+20A & L0   & 2205 & 19 $\pm$ 5 &    ---     & 18 $\pm$ 5 &    ---      & 18 $\pm$ 5 &    ---     & 20 $\pm$ 3 &    ---     &     ---   &19 $\pm$ 2\\
2MASS 0746+20B & L1.5 & 2060 & 33 $\pm$ 6 &    ---     & 32 $\pm$ 6 &    ---      & 32 $\pm$ 6 &    ---     & 34 $\pm$ 6 &    ---     &     ---   &33 $\pm$ 3 \\
2MASS 1426+16A & M8.5 & 2400 &     ---    & 6 $\pm$ 4  &    ---     & 6 $\pm$ 5   &    ---     & 7 $\pm$ 4  &    ---     &    ---     &     ---   &6 $\pm$ 3 \\
2MASS 1426+16B & L1   & 2240 &     ---    & 8 $\pm$ 7  &    ---     & 12$\pm$ 4   &    ---     & 11 $\pm$ 4 &    ---     &    ---     &     ---   &11 $\pm$ 3 \\                             
2MASS 1750+44A & M7.5 & 2200 &     ---    &    ---     &    ---     & 9 $\pm$ 3   &    ---     & 10 $\pm$ 4 &    ---     & 8 $\pm$ 4  &     ---   &9 $\pm$ 2 \\                                          
2MASS 1750+44B & L0   & 2020 &     ---    &    ---     &    ---     & 10 $\pm$ 4  &    ---     & 10 $\pm$ 5 &    ---     & 12 $\pm$ 4 &     ---   &11 $\pm$ 3 \\                                                            
2MASS 1847+55A & M7   & 2400 &     ---    & 3 $\pm$ 5  &    ---     & 4 $\pm$ 4   &    ---     &  9 $\pm$ 3 &    ---     & 8 $\pm$ 4  &     ---   &7 $\pm$ 2 \\                                                         
2MASS 1847+55B & M7.5 & 2100 &     ---    & 5 $\pm$ 4  &    ---     & 5 $\pm$ 5   &    ---     &  9 $\pm$ 4 &    ---     & 9 $\pm$ 4  &     ---   &7 $\pm$ 2 \\                                                               
2MASS 2140+16A & M8.5 & 2300 &     ---    & 13 $\pm$ 5 &    ---     & 12 $\pm$ 4  &    ---     & 11 $\pm$ 3 & 16 $\pm$ 4 &    ---     &     ---   &13 $\pm$ 2 \\                                                                 
2MASS 2140+16B & L2   & 2075 &     ---    & 42 $\pm$ 7 &    ---     & 37 $\pm$ 6  &    ---     & 34 $\pm$ 6 & 38 $\pm$ 7 &    ---     &     ---   &37 $\pm$ 3 \\                                                                 
2MASS 2206-20A & M8   & 2350 &     ---    & 20 $\pm$ 4 &    ---     & 18 $\pm$ 3  &    ---     & 19 $\pm$ 4 &    ---     & 16 $\pm$ 6 &     ---   &19 $\pm$ 2 \\                                                                   
2MASS 2206-20B & M8   & 2250 &     ---    & 22 $\pm$ 4 &    ---     & 20 $\pm$ 4  &    ---     & 21 $\pm$ 4 &    ---     & 18 $\pm$ 6 &     ---   &21 $\pm$ 2 \\                                                                 
GJ 569Ba       & M8.5 & 2000 &     ---    & 19 $\pm$ 3 &    ---     &    ---      &    ---     & 18 $\pm$ 3 &    ---     & 19 $\pm$ 3 &     ---   &19 $\pm$ 2 \\                                                            
GJ 569Bb       & M9   & 2000 &     ---    & 5 $\pm$ 5  &    ---     &    ---      &    ---     & 6 $\pm$ 5  &    ---     & 7 $\pm$ 4  &     ---   &6 $\pm$ 3 \\                                                           
HD 130948B     & L4   & 1840 &     ---    & 63 $\pm$ 8 &    ---     &    ---      &    ---     &     ---    &    ---     & 61 $\pm$ 7 &    62 $\pm$ 6   &62 $\pm$ 4 \\                                                      
HD 130948C     & L4   & 1790 &     ---    & 86 $\pm$ 8 &    ---     &    ---      &    ---     &     ---    &    ---     & 87 $\pm$ 11 &   84 $\pm$ 12  &86 $\pm$ 6 \\                                                      
LHS 2397aA     & M8   & 2180 &     ---    &    ---     & 14 $\pm$ 3 & 14 $\pm$ 3  & 15 $\pm$ 3 & 15 $\pm$ 3 & 15 $\pm$ 3 & 15 $\pm$ 3 &     ---   &15 $\pm$ 1 \\                                                                                 
LHS 2397aB     & L7.5 & 1350 &     ---    &    ---     & 10 $\pm$ 6 & 10 $\pm$ 6  & 14 $\pm$ 6 & 10 $\pm$ 8 & 10 $\pm$ 6 & 11 $\pm$ 8 &     ---   &11 $\pm$ 3 \\                                                                            
LP 349-25A     & M8   & 2200 & 56 $\pm$ 5 &    ---     & 50 $\pm$ 11&    ---      & 59 $\pm$ 7 & 55 $\pm$ 4 & 54 $\pm$ 4 &    ---     &     ---   &55 $\pm$ 2 \\                                                                         
LP 349-25B     & M9   & 2050 & 87 $\pm$ 6 &    ---     & 79 $\pm$ 11&    ---      & 76 $\pm$ 10& 81 $\pm$ 4 & 85 $\pm$ 5 &    ---     &     ---   &83 $\pm$ 3 \\                                                                         
LP 415-20A     & M7   & 2300 &     ---    &    ---     &    ---     &    ---      & 40 $\pm$ 6 &     ---    & 41 $\pm$ 7 &    ---     &     ---   &40 $\pm$ 5 \\                                                        
LP 415-20B     & M9.5 & 2000 &     ---    &    ---     &    ---     &    ---      & 36 $\pm$ 4 &     ---    & 40 $\pm$ 7 &    ---     &     ---   &37 $\pm$ 4 \\                                                          
\enddata
\tablenotetext{a}{From Konopacky et al. (2010).  To be
  conservative, our analysis
  assumed a $\pm$300 K temperature uncertainty for all objects}
\label{tab:vsini}
\end{deluxetable*}
%\clearpage
%\end{landscape}
%\end{singlespace}